\documentclass[twoside,twocolumn,english,secnumarabic,longbibliography,amsmath,nobibnotes,nofootinbib,aps,floatfix,superscriptaddress]{revtex4-1}
\usepackage{mathptmx}

\usepackage[latin9]{inputenc}
\usepackage[active]{srcltx}
\usepackage{color}
\usepackage{babel}
\usepackage{amsmath}
\usepackage{amssymb}
\usepackage{graphicx}
\usepackage{wasysym}
\usepackage[unicode=true,pdfusetitle,
 bookmarks=true,bookmarksnumbered=false,bookmarksopen=false,
 breaklinks=true,pdfborder={0 0 0},pdfborderstyle={},backref=false,colorlinks=true]
 {hyperref}
\hypersetup{
 linkcolor=blue,citecolor=blue,urlcolor=blue}
\usepackage{breakurl}

\makeatletter
\usepackage{float}
\usepackage{tikz}
\usetikzlibrary{calc,shapes,backgrounds}
\usetikzlibrary{datavisualization,patterns}
\usetikzlibrary{arrows.meta}
\usepackage{pslatex}
\usepackage{color}
\renewcommand{\textendash}{--}

\makeatother

\begin{document}

\title{Adaptive Density-Matrix Renormalization-Group study of the disordered
antiferromagnetic spin-1/2 Heisenberg chain}

\author{Alexander H. O. Wada\,\href{https://orcid.org/0000-0002-2425-3438}{\includegraphics[keepaspectratio,width=0.7em]{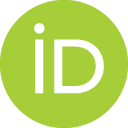}}}

\affiliation{Instituto de F\'{\i}sica de S\~ao Carlos, Universidade de S\~ao
Paulo, C.~P.~369, S\~ao Carlos, S\~ao Paulo 13560-970, Brazil}

\author{Jos\'e A. Hoyos\,\href{https://orcid.org/0000-0003-2752-2194}{\includegraphics[keepaspectratio,width=0.7em]{orcid.png}}}

\affiliation{Instituto de F\'{\i}sica de S\~ao Carlos, Universidade de S\~ao
Paulo, C.~P.~369, S\~ao Carlos, S\~ao Paulo 13560-970, Brazil}

\affiliation{Max Planck Institute for the Physics of Complex Systems, N\"othnitzer
Str. 38, 01187 Dresden, Germany}
\begin{abstract}
Using an adaptive strategy which enables the study of quenched disordered
system via the density-matrix renormalization-group method, we compute
the various ground-state spin-spin correlation measures of the spin-$1/2$
antiferromagnetic Heisenberg chain with random coupling constants,
namely, the mean values of the bulk and of the end-to-end correlations,
the typical value of the bulk correlations, and the distribution of
the bulk correlations. Our results are in agreement with the predictions
of the strong-disorder renormalization-group method. We do not find
any hint of logarithmic corrections either in the bulk average correlations,
which were recently reported by Shu \emph{et al}. {[}\hyperref{http://link.aps.org/doi/10.1103/PhysRevB.94.174442}{URL}{name}{Phys. Rev. B. {\bf 94}, 174442, (2016)}{]},
or in the end-to-end average correlations. We report the existence
of a logarithmic correction on the end-to-end correlations of the
clean chain. Finally, we have determined that the distribution of
the bulk correlations, when properly rescaled by an associated Lyapunov
exponent, is a narrow and universal (disorder-independent) probability
function.\\
\\
Published in \href{https://journals.aps.org/prb/abstract/10.1103/PhysRevB.105.104205}{Phys. Rev. B {\bf 105}, 104205 (2022)};
DOI: \href{https://doi.org/10.1103/PhysRevB.105.104205}{10.1103/PhysRevB.105.104205}
\end{abstract}

\maketitle

\section{Introduction}

One-dimensional random quantum systems display a rich plethora of
phenomena and are important theoretical laboratories for strongly
correlated quantum phenomena. A prominent phenomenon is the infinite-randomness
criticality (IRC)~\citep{Fisher_1992_RTFI}. Initially, it was thought
to be exclusive to one-dimensional systems. In recent decades, however,
it was found in many different and seemingly unrelated model systems.
To name a few, IRC governs the paramagnet\textendash ferromagnet transition
of the random transverse-field Ising model (in any dimension)~\citep{Fisher_2000,Kawashima_QM_RTFI,kovacs-igloi-prb11,vojta-hoyos-prl14}
and of the quenched disordered Hertz-Millis antiferromagnet~\citep{Hoyos_2007_DissipationQCP,vojta-kotabage-hoyos-prb09}
and the metal\textendash superconductor transition of rough thin films
and nanowires~\citep{delmaestro-rosenow-muller-sachdev-prl08,delmaestro-etal-prl10}.
IRC is also found in out-of-equilibrium situations such as Floquet
systems~\citep{berdanier-etal-pnas18} and reaction-diffusion classical
systems~\citep{Hooyberghs_2003_SDRG_absorbing} (for a review, see,
e.g., Refs.~\citealp{Igloi_2005_SDRGREview,Vojta_2006_RR,vojta-jltp-10,Igloi_2018_SDRGREview}).
Despite this plethora of theoretical situations, experimental checks
of IRC are still rare. Early hints come from quasi-one-dimensional
tetracyanoquinodimethan (TCNQ) compounds~\citep{bulaevskii,schegolev-pss72,azevedo-clark-prb77,tippie-clark-prb81}
{[}modeled by the Hamiltonian in Eq.~\eqref{eq:HXXZ}{]}, which initiated
this field of research. However, more accurate experiments and clearer
signatures are still desirable. In this context, precise knowledge
of the ground-state spin-spin correlation function (the main quantity
studied in this work) has great relevance as it dictates the behavior
of the structure factor at low temperatures~\citep{huse-structurefactor-PRL,CARandomSpinChain_Hoyos2007},
which is experimentally accessible via neutron-scattering experiments.
Finally, it is worth noting that, more recently, strong experimental
evidence of infinite-randomness criticality was reported in itinerant
magnets~\citep{guo-etal-prl08,westercamp-etal-prl09,ubaidkassis-vojta-schroeder-prl10}
(for a review, see Ref.~\citealp{vojta-jltp-10}) and in thin superconducting
films~\citep{xing-etal-science15}.

A paradigmatic model exhibiting IRC is the random antiferromagnetic
(AF) spin-$1/2$ XXZ chain 
\begin{equation}
H=\sum_{i}J_{i}\left(S_{i}^{x}S_{i+1}^{x}+S_{i}^{y}S_{i+1}^{y}+\Delta S_{i}^{z}S_{i+1}^{z}\right),\label{eq:HXXZ}
\end{equation}
 where $\mathbf{S}_{i}$ are the usual spin-1/2 operators associated
with site $i$, the antiferromagnetic coupling constants $J_{i}>0$
are independent and identically distributed random variables drawn
from a distribution $P_{D}(J)$ {[}with $D$ parametrizing the disorder
strength; see, for definiteness, Eq.~\eqref{eq:P}{]}, and $\Delta$
is the anisotropy parameter. It is now well accepted that, for $-\frac{1}{2}<\Delta\leq1$,
the chain is critical and governed by an infinite-randomness fixed
point where the arithmetic and geometric means (henceforth referred
to as mean and typical values, respectively) of the spin-spin correlation
function {[}$C_{i}^{\alpha}(r)=\left\langle S_{i}^{\alpha}S_{i+r}^{\alpha}\right\rangle $,
with $\left\langle \cdots\right\rangle $ denoting the ground-state
average{]} behave quite differently. 

In the thermodynamic limit and for spins sufficiently far from each
other, the mean value is 
\begin{equation}
\overline{C_{i}^{\alpha}}(r)=\frac{\left(-1\right)^{r}}{12r^{\eta}}\begin{cases}
c_{\alpha,\text{o}}, & \mbox{if }r\mbox{ is odd},\\
c_{\alpha,\text{e}}, & \mbox{otherwise},
\end{cases}\label{eq:C-mean}
\end{equation}
 with $\overline{\cdots}$ denoting the arithmetic average over the
disorder configurations. The exponent $\eta=2$ is universal {[}i.e.,
does not depend on the details of $P_{D}(J)${]}, isotropic (i.e.,
$\alpha$ independent), and $\Delta$ independent~\citep{Fisher_1994_RTFI}
due to an enhancement in the ground-state symmetry from SO($N$)$\rightarrow$SU($N$)
(here, $N=2$), a generic feature of SO($N$)-symmetric AF random
spin chains~\citep{quito-etal-epjb20,quito-etal-prb19}. The numerical
prefactors $c_{\alpha,\text{o,e}}$, on the other hand, are nonuniversal
(i.e., disorder dependent), anisotropic (i.e., $\alpha$ dependent),
and $\Delta$ dependent. Surprisingly, it was conjectured that $c_{\alpha,\text{o}}-c_{\alpha,\text{e}}=1$
is universal if $\alpha$ is a symmetry axis, i.e., for $\alpha=z$,
and for any $\alpha$ when $\Delta=1$~\citep{CARandomSpinChain_Hoyos2007}.

The typical value of the spin-spin correlation function, 
\begin{equation}
C_{\text{typ}}^{\alpha}\left(r\right)\equiv\exp\overline{\ln\left|\left\langle S_{i}^{\alpha}S_{i+r}^{\alpha}\right\rangle \right|}\approx c_{\alpha,D}\exp\left[-A_{\alpha}\times\left(r\gamma_{D}\right)^{\psi}\right],\label{eq:C-typ}
\end{equation}
 behaves quite differently. It decays stretched exponentially with
universal and isotropic tunneling exponent $\psi=\frac{1}{2}$~\citep{Fisher_1994_RTFI}.
The numerical prefactor $A_{\alpha}$ is universal and anisotropic,
and the Lyapunov exponent $\gamma_{D}$ is nonuniversal, isotropic,
and $\Delta$ dependent. For the free-fermionic case $\left(\Delta=0\right)$,
a single-parameter theory~\citep{mard-etal-prb14,Getelina_CA} predicts
that $\gamma_{D}=8\pi^{-1}\text{var}\left(\ln J\right)$ {[}where
$\text{var}\left(x\right)=\overline{x^{2}}-\overline{x}^{2}$ is the
variance{]}. For the generic case $\left(-\frac{1}{2}<\Delta\leq1\right)$,
however, $\gamma_{D}\propto\left[\text{var}\left(\ln J\right)\right]^{\frac{1}{3-2K}}$
with $2K=\left[1-\pi^{-1}\arccos\left(\Delta\right)\right]^{-1}$.\footnote{According to standard field-theory methods~\citep{SineGordenCorr_Giamarchi1989,DotyFisher_1992_QuenchedDisorderSpinHalfXXZ},
the Lyapunov exponent is $\gamma_{D}\propto\left[\text{var}\left(J\right)\right]^{\frac{1}{3-2K}}$.
While this is accurate for $D\ll1$, it was numerically shown that
$\gamma_{D}\propto\left[\text{var}\left(\ln J\right)\right]^{\frac{1}{3-2K}}$
is a much better choice for any $D$~\citep{Florencie_XXZCrossLength}.} 

Results \eqref{eq:C-mean} and \eqref{eq:C-typ} stem from the fact
that the ground state is a random singlet which is captured by the
strong-disorder renormalization-group (SDRG) method and, supposedly,
are asymptotic exact~\citep{Fisher_1994_RTFI}. It is worth mentioning
that, at the free-fermion point $\Delta=0$~\citep{henelius-girvin,laflorencie-correlacao-PRL,CARandomSpinChain_Hoyos2007,Getelina_CA}
and at the isotropic Heisenberg point $\Delta=1$~\citep{Florencie_XXZCrossLength,Bonesteel_valencebondfluctuations,Romer_TTN_SDRG,Goldsborough_renormalization4disorderedsystems},
these results (among other SDRG predictions) have been confirmed with
increasing numerical precision over the years (see Refs.~\citealp{Igloi_2005_SDRGREview,Igloi_2018_SDRGREview}
and references therein). Interestingly, however, a recent ground-state
quantum Monte Carlo study found a logarithmic factor in the mean correlation
function~\citep{XXXLogCorr_Sandvik2016} at the isotropic point $\Delta=1$.
Namely, result \eqref{eq:C-mean} is corrected to 
\begin{equation}
\overline{C^{\alpha}}(r)\sim\ln^{\sigma_{s}}r/r^{2},\label{eq:C-log-corrections}
\end{equation}
 with $0.3\apprle\sigma_{s}\apprle0.7$. It is certainly desirable
to understand the origin of this logarithmic correction, which is
not predicted by the SDRG method.\footnote{Recently, the subleading corrections to \eqref{eq:C-mean} in the
SDRG framework were obtained, and no hint of logarithmic corrections
was found~\citep{juhasz-prb21}.} For the homogeneous (clean) system at the Heisenberg point $\Delta=1$~\citep{Affleck_1989CentralCharge,SineGordenCorr_Giamarchi1989,singh-etal-prb89,hallberg-etal-prb95,Affleck_1998,LowEH-XXZ_Lukyanov1998,CorAmpXXZ-DMRG_Hikihara1998,sandvik-AIPcp-10}
and for the dirty system at $\Delta=-\frac{1}{2}$ (where disorder
is perturbatively irrelevant)~\citep{ristivojevic-etal-npb12}, logarithmic
factors due to marginally irrelevant operators have been reported.
Which marginal operator, if any, endows the logarithmic factor to
$\overline{C^{\alpha}}$? Does the typical value also acquire a similar
correction? Unfortunately, conventional perturbative field-theoretical
methods cannot be applied at $\Delta=1$ due to runaway flow of the
disorder strength.

Furthermore, it is interesting to ponder the consequences of a possible
logarithmic factor to the correlation function. Assuming that the
resulting random singlet ground state is localized, i.e., the typical
correlation is stretched exponentially small (regardless of logarithmic
corrections), it is then possible to use the methods of Ref.~\citealp{CARandomSpinChain_Hoyos2007}
to relate $\overline{C^{\alpha}}(r)$ to the von Neumann entanglement
entropy 
\begin{equation}
{\cal S}_{l}=-\textrm{Tr}\rho_{A}\ln\rho_{A},\label{eq:EE}
\end{equation}
 where $\rho_{A}$ is the reduced density matrix of subsystem $A$
(of length $l$) obtained by tracing the degrees of freedom of the
complementary subsystem $B$. To leading order in $l$, they are related
via ${\cal S}_{l}=-8\ln2\sum_{r=1}^{l}\overline{C^{\alpha}}(r)r\sim\left(\ln l\right)^{1+\sigma_{s}}$.
Thus, this would be an interesting violation of the area law if $\sigma_{s}\neq0$.

It is thus desirable to confirm the existence of the logarithmic factor
found in Ref.~\citealp{XXXLogCorr_Sandvik2016}. Therefore, we study
the spin-spin correlation function of the random AF spin-1/2 Heisenberg
chain {[}Eq.~\eqref{eq:HXXZ} with $\Delta=1${]} using the adaptive
density-matrix renormalization-group (aDMRG) method, which is a recently
introduced unbiased method for strongly disordered systems.

The remainder of this paper is organized as follows. In Sec.~\eqref{sec:Clean-and-disordered}
we define the coupling constant distribution $P_{D}(J)$ and review
the employed aDMRG method. In Sec.~\eqref{sec:numerics} we apply
the DMRG method to the clean chain, and we show that both the bulk
and the end-to-end correlation exhibit logarithmic factors. We then
apply the aDMRG method to the disordered case and study the effects
of disorder on the bulk mean and typical values of the correlation,
the end-to-end correlation, and the distribution of the correlations.
In all cases, our data are compatible with the absence of logarithmic
factors. Finally, we summarize and discuss our results in Sec.~\eqref{sec:Conclusion}.

\section{\label{sec:Clean-and-disordered}Disorder description and method }

We study the ground-state spin-spin correlation function of the random
AF spin-1/2 Heisenberg chain. The model Hamiltonian is given by Eq.~\eqref{eq:HXXZ}
with $\Delta=1$. The coupling constants $0<J_{i}<1$ are uncorrelated
random variables drawn from the probability distribution

\begin{equation}
P_{D}(J)=D^{-1}J^{1/D-1},\label{eq:P}
\end{equation}
 where the disorder strength is parametrized by $D$: $\overline{\ln^{2}J}-\overline{\ln J}^{2}=D^{2}$.
$D=0$ is the clean chain, while $D\rightarrow\infty$ is the infinitely
disordered case.

In Sec.~\eqref{subsec:dirty}, all the data are averaged over $2\times10^{4}$
distinct disorder configurations of coupling constants $\left\{ J_{i}\right\} $.

How the efficiency of the DMRG method diminishes when dealing with
systems governed by infinite-randomness physics is notorious~\citep{juozapavicius-etal-prb97,Wenzel_2002,laflorencie-correlacao-PRL,ruggiero-etal-prb16}.
The reason is due to a disorder-induced rough energy landscape with
nearly degenerate local minima. The standard DMRG method then gets
stuck in an excited/metastable state. As a result, the method fails
to capture the rare spin pairs (or clusters) that are largely separated
but highly entangled. Although rare, they are responsible for the
leading contribution to the mean value of the spin-spin correlations.

In order to circumvent this problem, we employ the recently introduced
adaptive aDMRG method~\citep{Xavier_ADMRG} to obtain the ground-state
spin-spin correlation function. The idea is to apply the standard
DMRG method to a clean or nearly clean system (where it works efficiently
well) in order to obtain a good representation of the ground state
$\left|\psi_{D_{0}}\right\rangle $ and then modify it adiabatically
by increasing the disorder strength $D$ in small steps $D\rightarrow D+\delta D$.
Precisely, (i) we start with a disorder configuration $\left\{ J_{i}\right\} $
drawn from \eqref{eq:P} with $D=D_{0}\ll1$. The standard DMRG method
is then applied, and $\left|\psi_{D}\right\rangle $ is obtained (after
convergence). The next step is to (ii.a) increase the disorder strength
to $D+\delta D$ while keeping the disorder configuration fixed; that
is, we simply make the transformation $J_{i}\rightarrow J_{i}^{1+\frac{\delta D}{D}}$.
(ii.b) Using the previously found ground state $\left|\psi_{D}\right\rangle $
as an input, the standard DMRG method is applied again, from which
$\left|\psi_{D+\delta D}\right\rangle $ is obtained. (iii) Step (ii)
is iterated until the desired disorder strength $D$ is reached.

We have used $D_{0}=\delta D=1/16$ (the nearly clean system). Our
DMRG code is implemented using the ITensor Library~\citep{itensor}
on chains of $L$ spins with open boundary conditions. In each DMRG
application we kept up to $N=400$ states, which is enough to keep
the truncation error below $T_{\text{err}}\sim10^{-10}$. To ensure
convergence, we used $20$ sweeps for the initial state $\left|\psi_{D_{0}}\right\rangle $
and $\delta_{\text{sweeps}}=4$ sweeps when increasing the disorder
strength, i.e., when going from $\left|\psi_{D}\right\rangle \rightarrow\left|\psi_{D+\delta D}\right\rangle $.

\section{Numerical results\label{sec:numerics}}

In this section, we report our numerical results using the aDMRG method
on the various spin-spin correlation functions studied: the mean and
typical values for the bulk, the mean end-to-end correlations, and
the distribution of the bulk correlations. They are studied for the
cases of homogeneous and randomly disordered chains. Finally, we have
studied only chains with open boundary conditions.

\subsection{The homogeneous AF Heisenberg chain\label{subsec:clean}}

Due to the open boundary conditions, the system is not translation
invariant, and therefore, we average over the various spin pairs of
the same size; that is, the bulk correlation function is defined as

\begin{equation}
C(r)=\frac{\sum_{i=L/4}^{3L/4-r}\langle\mathbf{S}_{i}\cdot\mathbf{S}_{i+r}\rangle}{L/2-r},\label{eq:C}
\end{equation}
 where, in order to reduce the finite-size effects, we have excluded
the $L/4$ spins closest to the open boundaries.

\begin{figure}[tb]
\begin{centering}
\includegraphics[clip,width=0.99\columnwidth]{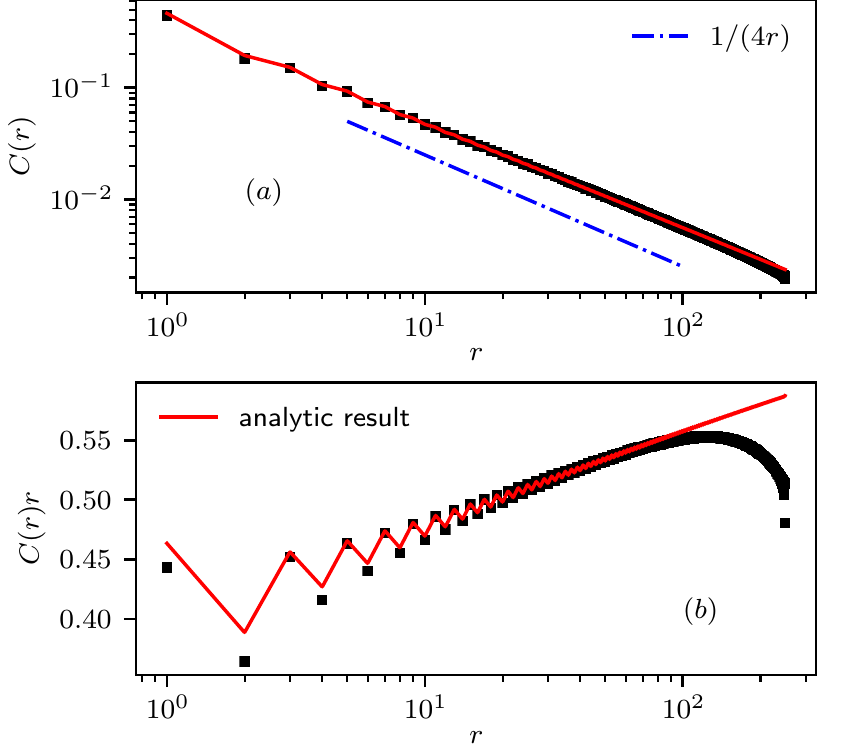}
\par\end{centering}
\caption{The spin-spin correlation function \eqref{eq:C} (black squares) as
a function of the spin separation $r$ for a chain of $L=500$ sites
with open boundary conditions. The blue dashed line in (a) is the
simple power-law decay $\sim r^{-1}$ . In (b) we plot $C(r)r$ to
emphasize the logarithmic correction. The solid red line is the best
fit to the analytical expectation \eqref{eq:LukC} in both panels.\label{fig:Cclean} }
\end{figure}

We plot in Fig.~\hyperref[fig:Cclean]{\ref{fig:Cclean}(a)} $C(r)$
as a function of the spin-spin separation $r$ for a chain of $L=500$
spins (black squares). In Fig.~\hyperref[fig:Cclean]{\ref{fig:Cclean}(b)}
we replot the same data multiplied by $r$ in order to highlight the
logarithmic correction.

The leading terms of the correlation function $\left\langle \mathbf{S}_{i}\cdot\mathbf{S}_{i+r}\right\rangle $
in the regime $1\ll r\ll L$ are known to be~\citep{LowEH-XXZ_Lukyanov1998,CorAmpXXZ-DMRG_Hikihara1998} 

\begin{equation}
\left\langle \mathbf{S}_{i}\cdot\mathbf{S}_{i+r}\right\rangle =a\frac{A(r)}{r}+b\frac{(-1)^{r}}{r^{2}},\label{eq:LukC}
\end{equation}
 with $a=1$ and $b$ being an unknown constant (both of which we
take as fitting parameters for our numerical data) and the function 

\begin{equation}
A(r)=\frac{3}{\sqrt{8\pi^{3}g}}\left(1-\frac{3}{16}g^{2}+\frac{156\zeta(3)-73}{384}g^{3}+{\cal O}\left(g^{4}\right)\right),\label{eq:A}
\end{equation}
 where $\zeta\left(s\right)$ is the Riemann zeta function and $g\equiv g(r)$
is obtained from 

\begin{equation}
g^{-1}+\frac{1}{2}\ln g=\ln\left(2\sqrt{2\pi}e^{\gamma+1}r\right),
\end{equation}
where $\gamma$ is the Euler constant. In both panels of Fig.~\ref{fig:Cclean}
we fit our numerical data to the analytical expectation \eqref{eq:LukC}
and find that $a=0.983(1)$ and $b=-0.452(9)$ (red solid line).\footnote{If instead of \eqref{eq:C-mean} one defines $C(r)=\langle\mathbf{S}_{(L-r)/2}\cdot\mathbf{S}_{(L+r)/2}\rangle$
for $r$ even, and 2$C(r)=\langle\mathbf{S}_{(L-r+1)/2}\cdot\mathbf{S}_{(L+r+1)/2}\rangle+\langle\mathbf{S}_{(L-r-1)/2}\cdot\mathbf{S}_{(L+r-1)/2}\rangle$
for $r$ odd, only the last digit of the fitting parameters to $a$
and $b$ changes.} Therefore, we confirm that, to leading order, $\left\langle \mathbf{S}_{i}\cdot\mathbf{S}_{i+r}\right\rangle \sim\sqrt{\ln r}/r$.

We now study the end-to-end correlation function. For large system
sizes ($L\gg1$), we expect that 
\begin{equation}
C_{1,L}=\left\langle \mathbf{S}_{1}\cdot\mathbf{S}_{L}\right\rangle =c\frac{[\ln(L/L_{0})]^{\theta}}{L^{\eta_{s}}},\label{eq:CL}
\end{equation}
 where the surface correlation function exponent $\eta_{s}=2x_{s}=2$~\citep{CI-Surface_Cardy1984,SurfaceXXZ-AT-P_Alcaraz1987}.
For the same reason as in the bulk correlations, we expect a logarithmic
factor. To the best of our knowledge, however, the exponent $\theta$
is unknown. 

\begin{figure}[tb]
\begin{centering}
\includegraphics[clip,width=0.99\columnwidth]{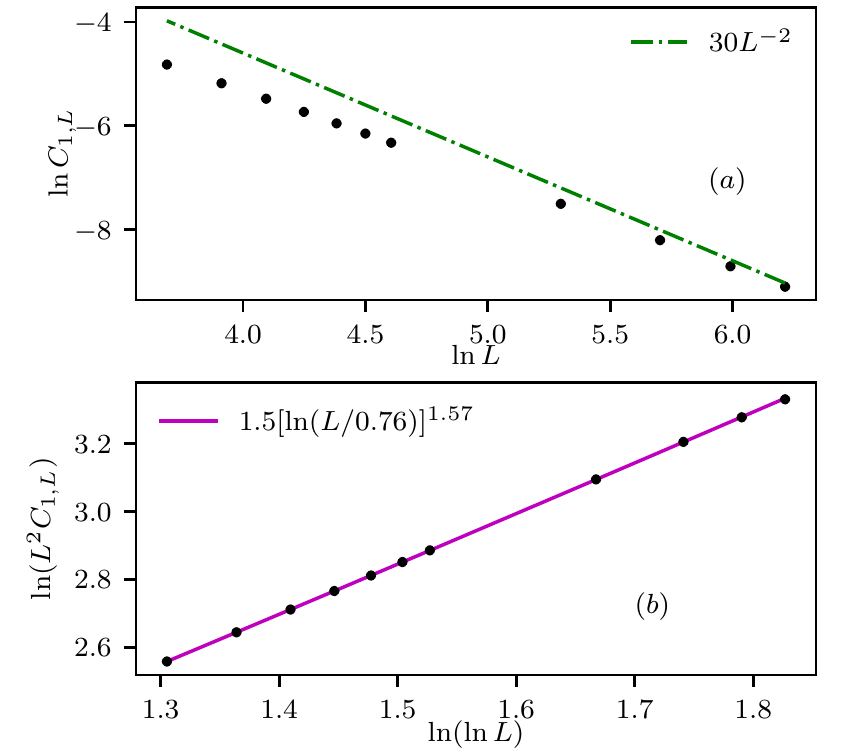}
\par\end{centering}
\caption{(a) The end-to-end correlation $C_{1,L}=\left\langle \mathbf{S}_{1}\cdot\mathbf{S}_{L}\right\rangle $
for system sizes ranging from $L=40$ to $L=500$. (b) Same data as
in (a) with $C_{1,L}$ multiplied by $L^{2}$ in order to highlight
the logarithmic prefactor {[}see Eq.~\eqref{eq:CL}{]}. \label{fig:CL} }
\end{figure}

Figure \hyperref[fig:CL]{\ref{fig:CL}(a)} shows $C_{1,L}$ for system
sizes ranging from $L=40$ up to $500$. Clearly, $C_{1,L}$ does
not decay as a simple power law $\sim L^{-2}$. In Fig.~ \hyperref[fig:CL]{\ref{fig:CL}(b)}
we plot $C_{1,L}L^{\eta_{s}}$, from which we fit Eq.~\eqref{eq:CL}
to our data taking $\theta$, $c$, and $L_{0}$ as fitting parameters.
We obtain $L_{0}=0.7$(2), $c=1.5(3)$, and $\theta=1.5(2)$.\footnote{The error in the last digit of the fitting parameters is obtained
by removing the first tree data points (smallest $L$'s) from the
fit.} Evidently, the value of the exponent $\theta\approx1.5$ should be
interpreted only as an effective exponent since we are not performing
a thorough finite-size study.

\subsection{The disordered AF Heisenberg chain\label{subsec:dirty}}

In this section, we report our main results on the ground-state correlation
function of the AF disordered Heisenberg chain {[}$\Delta=1$ in the
Hamiltonian \eqref{eq:HXXZ}{]} using the aDMRG method.

\begin{figure}[tb]
\begin{centering}
\includegraphics[clip,width=0.99\columnwidth]{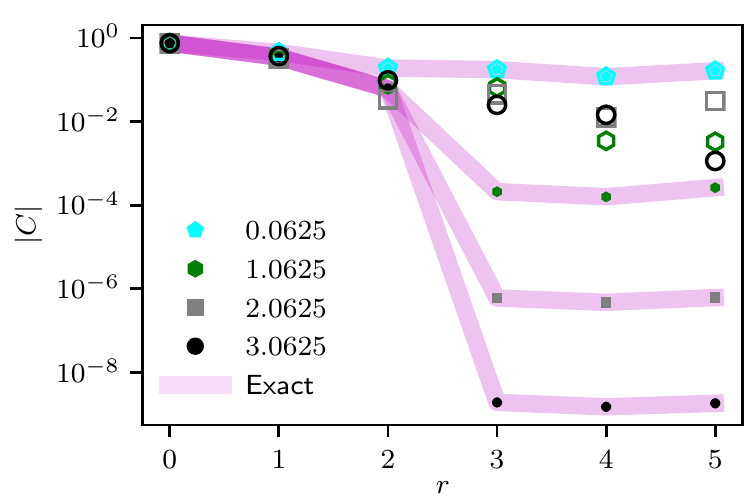}
\par\end{centering}
\caption{The ground-state spin-spin correlation \eqref{eq:C} for a single
disorder realization $\left\{ J_{i}\right\} $ drawn from \eqref{eq:P}
for various disorder strengths $D$ (increased in steps by the aDMRG
method) and system size $L=10$. The thick solid curve, open symbols,
and solid symbols are, respectively, the values obtained using the
exact diagonalization, the standard DMRG, and aDMRG methods.\label{fig:benchmark}}
\end{figure}

As a benchmark, we start by computing the correlation function $C$
{[}as defined in Eq.~\eqref{eq:C}{]} for a single disorder realization
of coupling constants $\{J_{i}\}$ using the exact diagonalization,
the standard DRMG, and the aDMRG methods for a chain of $L=10$ spins.
As shown in Fig.~\eqref{fig:benchmark}, the standard DMRG method
fails to reproduce the exact values, while the aDMRG method reproduces
the exact ones within a relative error smaller than $10^{-3}$. We
have repeated this benchmark for dozens of other disorder realizations
and have obtained the same result.\footnote{For further comparison between these methods, we refer the reader
to Ref.~\citealp{Xavier_ADMRG}.} 

\begin{figure}[t]
\begin{centering}
\includegraphics[clip,width=0.99\columnwidth]{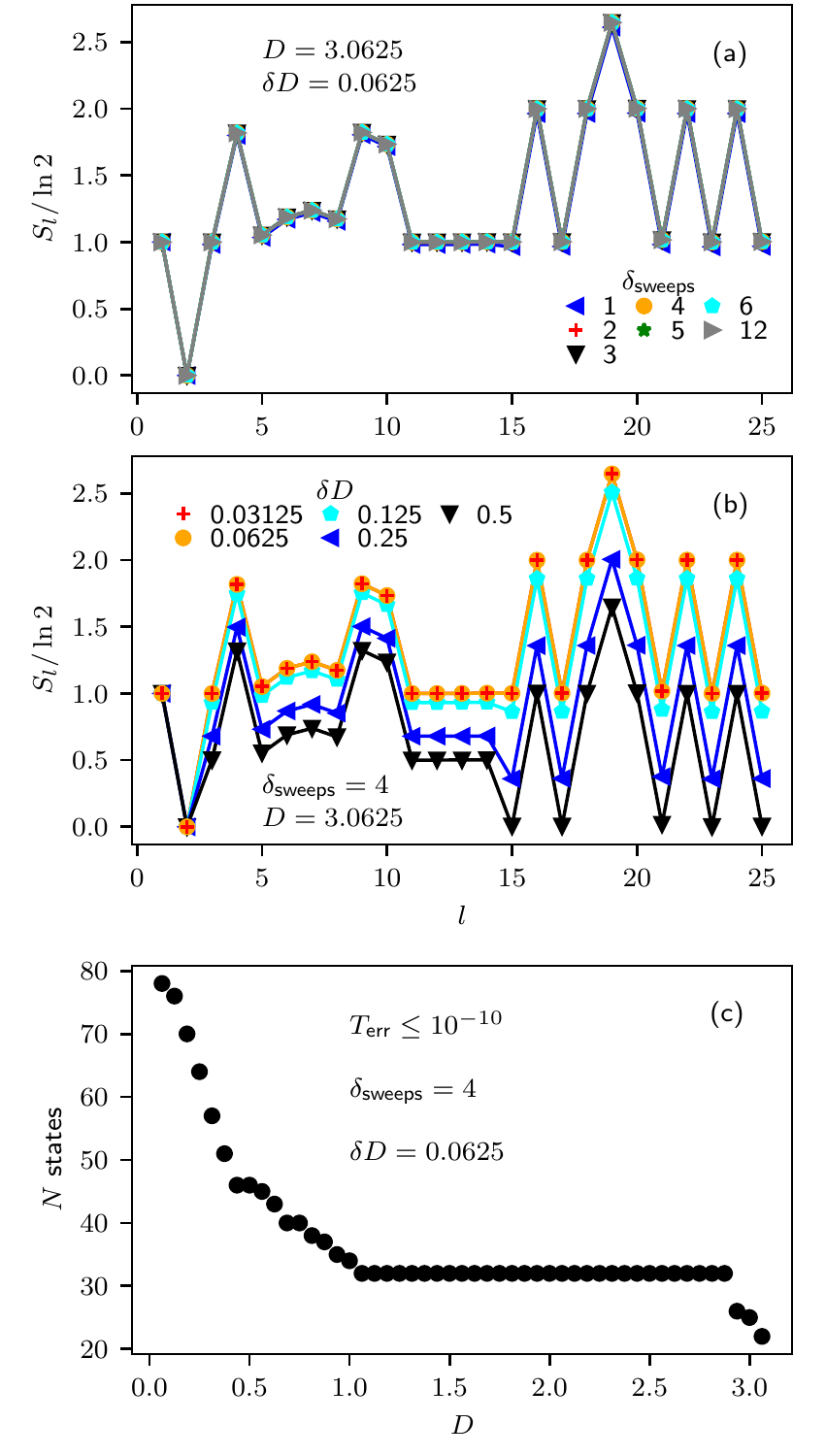}
\par\end{centering}
\caption{(a) and (b) The entanglement entropy ${\cal S}_{l}$ as a function
of the system size $l$ {[}see Eq.~\eqref{eq:EE}{]} for a single
disorder realization of a chain of $50$ sites long and disorder parameter
strength $D=3.0625$. The subsystem $A$ consists of all spins from
sites $1$ to $l$. In (a), ${\cal S}_{l}$ is shown for different
numbers of DMRG sweeps $\delta_{\text{sweeps}}$ used in the adaptive
strategy while the disorder increment $\delta D$ is kept fix. In
(b), ${\cal S}_{l}$ is shown for different values of $\delta D$
and fix $\delta_{\text{sweep}}$. (c) The total number of states $N$
needed to ensure that the DMRG truncation error is not larger than
$T_{\text{err}}=10^{-10}$ as the disorder strength is increased.
\label{fig:convergence}}
\end{figure}

In order to illustrate the convergence of the adaptive strategy, we
plot in Fig.~\ref{fig:convergence} an additional analysis with respect
to the (a) number of DMRG sweeps $\delta_{\text{sweeps}}$ necessary
for convergence when increasing the disorder strength from $D\rightarrow D+\delta D$,
(b) the disorder strength increment $\delta D$, and (c) the number
of states needed to keep the DMRG error truncation $T_{\text{err}}$
below a certain threshold. We then compute the entanglement entropy
$S_{l}$ \eqref{eq:EE} as a function of the subsystem size $l$.
Here, we show only a typical disorder realization of a chain $L=50$
sites long with final disorder strength $D=3.0625$, but we have checked
the same quantitative results for other chain sizes. In Fig.~\hyperref[fig:convergence]{\ref{fig:convergence}(a)},
we see that $\delta_{\text{sweeps}}=2$ or $3$ is already enough
to ensure convergence of ${\cal S}_{l}$ (and, presumably, of the
state $\left|\psi_{D}\right\rangle $). In Fig.~\hyperref[fig:convergence]{\ref{fig:convergence}(b)},
we see that a small disorder parameter increment $\delta D\lessapprox2^{-4}$
is necessary in order to obtain convergence. We notice that increasing
the number of intermediate sweeps $\delta_{\text{sweeps}}$ does not
improve convergence for larger $\delta D$'s. Finally, we plot in
Fig.~\hyperref[fig:convergence]{\ref{fig:convergence}(c)} the total
number of states $N$ required to keep the DMRG truncation error below
$T_{\text{err}}=10^{-10}$ as the disorder strength $D$ is increased
along the adaptive strategy. Clearly, the disorder gets larger, fewer
states are necessary. We report that the entanglement entropy ${\cal S}_{l}$
converges less rapidly than the correlation function $C$; that is,
if the parameters used are enough to ensure the convergence of ${\cal S}_{l}$,
then $C$ is also converged.

\begin{figure}[tb]
\begin{centering}
\includegraphics[clip,width=0.99\columnwidth]{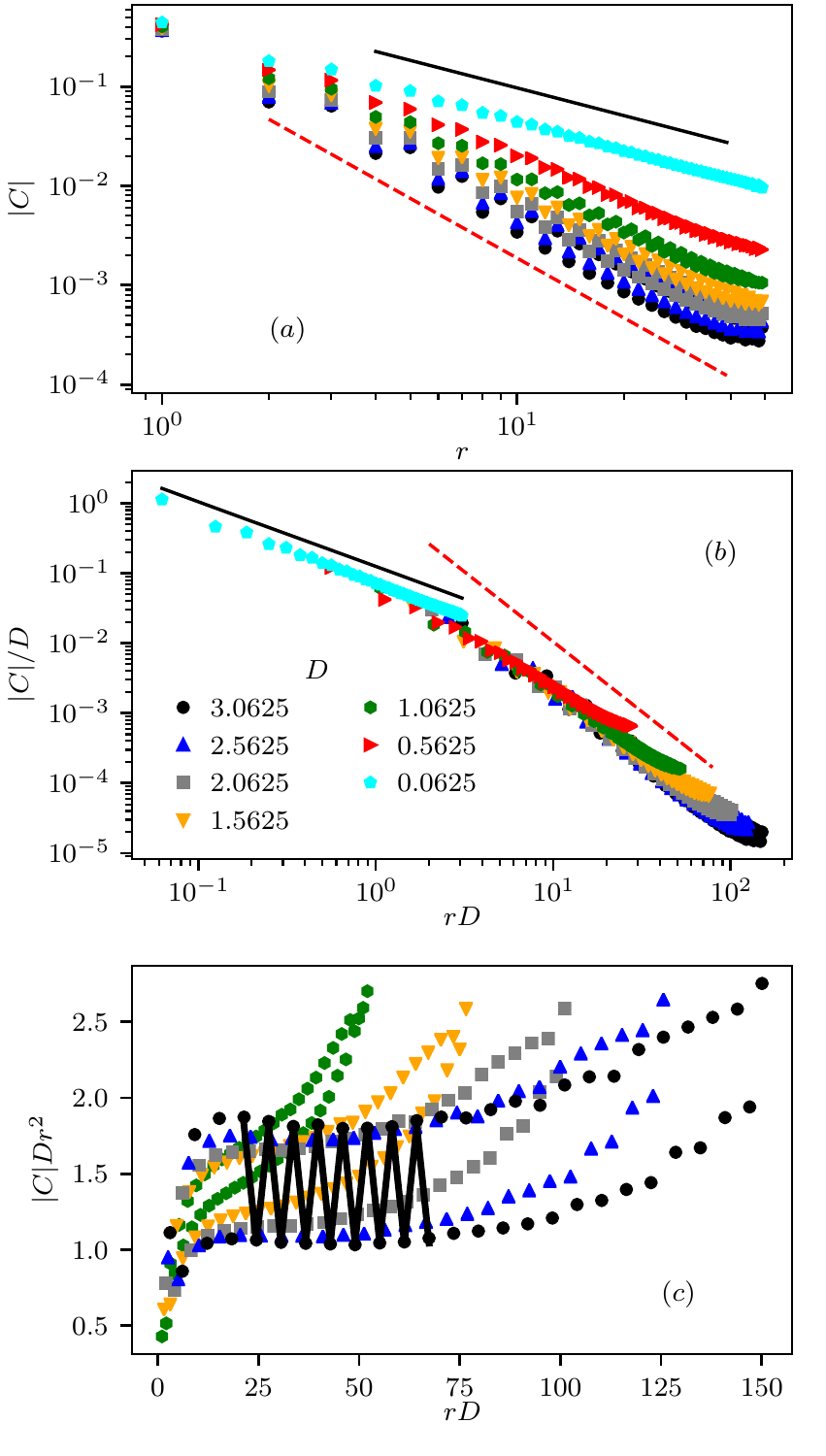}
\par\end{centering}
\caption{The mean value of the correlation function \eqref{eq:C} of the AF
Heisenberg chain for $L=100$ as a function of the spin-spin separation
$r$ for various disorder strengths $D$. (a) shows the bare value
of $\left|\overline{C}\right|$ as a function of the spin-spin separation
$r$. In (b), we plot $\left|\overline{C}\right|/\gamma_{D}$ as a
function of $\gamma_{D}r$ in order to obtain a data collapse {[}see
arguments around Eq.~\eqref{eq:Lyapunov}{]}. The black solid line
is proportional to the clean correlation \eqref{eq:LukC}, and the
red dashed line is proportional to $r^{-2}$. The data in (b) are
replotted in (c) with the vertical axis multiplied by $\left(\gamma_{D}r\right)^{2}$.
The solid line is a fit to Eq.~\eqref{eq:C-mean}. \label{fig:Crand}}
\end{figure}

Now we turn to the main results of this work. The first one is on
the mean value of the bulk correlations $\overline{C}(r)$ {[}as defined
in \eqref{eq:C}{]} for chains of $L=100$ spins and various disorder
strengths $D$, shown in Fig.~\ref{fig:Crand}. In Fig.~\hyperref[fig:Crand]{\ref{fig:Crand}(a)}
we can see that $\overline{C}$ crosses over from the clean behavior
$C_{\text{clean}}\sim(\sqrt{\ln r})/r$ (black solid line) to the
disordered one $\overline{C}\sim r^{-2}$ (red dashed line) with increasing
disorder strength $D$, as expected.

In order to obtain a data collapse, we now follow the reasoning of
Ref.~\citealp{Getelina_CA}. The first step is to relate the clean-dirty
crossover length $\xi_{D}$ to the multiplicative prefactor $c_{D}$
of the correlation function $\overline{C}=c_{D}r^{-2}$. This is accomplished
by assuming a sharp crossover at $r=\xi_{D}$, i.e., $C_{\text{clean}}=A(\xi_{D})/\xi_{D}=\overline{C}=c_{D}/\xi_{D}^{2}$,
and thus, $c_{D}\sim\xi_{D}A(\xi_{D})$, with $A(r)$ defined in \eqref{eq:A}.
The second step is to rescale the spin-spin separation $r$ in terms
of $\xi_{D}$ (i.e., $r\rightarrow r/\xi_{D}$) and to rescale $\overline{C}$
accordingly. Thus, $\overline{C}\xi_{D}/A(\xi_{D})\sim\left(r/\xi_{D}\right)^{-2}$.
The third step is to relate $\overline{C}$ to the disorder strength
$D$. As explained in the \hyperref[sec:intro]{Introduction}, the
associated Lyapunov exponent is 
\begin{equation}
\gamma_{D}\equiv\left[\text{var}\left(\ln J\right)\right]^{\frac{1}{3-2K}}.\label{eq:Lyapunov}
\end{equation}
 with $2K=\left[1-\pi^{-1}\arccos\left(\Delta\right)\right]^{-1}$~\citep{DotyFisher_1992_QuenchedDisorderSpinHalfXXZ},
and we ignore any possible multiplicative prefactor. As $\xi_{D}\propto\gamma_{D}^{-1}$,
then, $\xi_{D}\propto D^{-1}$ for $\Delta=1$. In order to proceed,
we need to take a final step: we assume that $A(\xi_{D})\approx1$
for $D>1$. This is justified after we verify that $\xi_{D}$ is of
order unity in the data in Fig.~\hyperref[fig:Crand]{\ref{fig:Crand}(a)}.
While $A(\xi_{D})$ cannot be dropped for $D<1$, for our purposes
we need the correct scaling function only in the strong-disorder regime.
Thus, we plot $\left|\overline{C}\right|/D$ vs $rD$ in Fig.~\hyperref[fig:Crand]{\ref{fig:Crand}(b)}.
The data collapse reasonably well apart from the deviations due to
finite-size effects and $D$-dependent corrections to $A(\xi_{D})$
(or $c_{D}$) in the $D<1$ regime.\footnote{Reference \citealp{Getelina_CA} showed that these corrections, although
smaller, exist even in the XX chain, where there are no logarithmic
corrections to the clean correlation function.} Finally, in Fig.~\hyperref[fig:Crand]{\ref{fig:Crand}(c)} we plot
$\overline{C}Dr^{2}$ as a function of $rD$, which should be compared
to the clean case in Fig.~\hyperref[fig:Cclean]{\ref{fig:Cclean}(b)}.
The increasing of the plateau for the largest values of $D$ suggests
the nonexistence of the logarithmic factor. Notice that this is reached
only for the largest values of $D$. For $D\approx2$, the plateau
seems like a shoulder and is strongly affected by the finite-size
corrections. As reported in Ref.~\citealp{Getelina_CA}, this can
mimic logarithmic factors. We remark that the numerical observation
of this plateau is not a simple task to accomplish even in the free-fermion
case $\Delta=0$ with periodic boundary conditions~\citep{Getelina_CA}.

\begin{figure}[tb]
\begin{centering}
\includegraphics[clip,width=0.99\columnwidth]{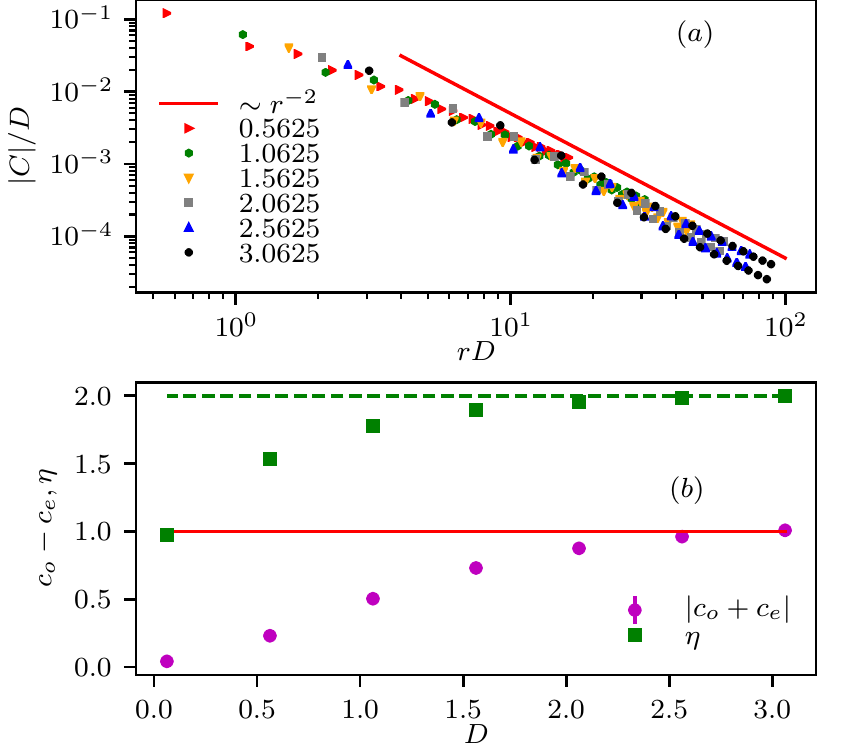}
\par\end{centering}
\caption{(a) Replot of the data in Fig.~\hyperref[fig:Cclean]{\ref{fig:Crand}(b)}
excluding the data points which do not follow the data collapse. (b)
The exponent $\eta$ and the difference $c_{\text{o}}-c_{\text{e}}$
obtained from the best fit of Eq.~\eqref{eq:C-mean} to the data
in (a) (see text). \label{fig:eta-c}}
\end{figure}

We now extract the value of the exponent $\eta$ and the difference
$c_{\text{o}}-c_{\text{e}}$ between the numerical prefactors. We
then analyze the data in Fig.~\hyperref[fig:Crand]{\ref{fig:Crand}(b)}
excluding the points which, due to strong finite-size effects, are
out of the data collapse. The resulting data points are replotted
Fig.~\hyperref[fig:eta-c]{\ref{fig:eta-c}(a)} from which we fit
Eq.~\eqref{eq:C-mean} to each data set using $\eta$, $c_{\text{o}}$,
and $c_{\text{e}}$ as fitting parameters. The corresponding values
are plotted in Fig.~\hyperref[fig:eta-c]{\ref{fig:eta-c}(b)} as
a function of the disorder strength $D$. For small values of $D$,
$\eta$ is simply an effective exponent due to the large associated
crossover length. As $D$ increases, the crossover length shortens,
and the effective exponent $\eta$ approaches the expected value $\eta=2$.
We observe analogous behavior for the difference $c_{\text{o}}-c_{\text{e}}$.
The fitted values for $D=3.0625$ are $\eta=1.99(2)$ and $c_{\text{o}}-c_{\text{e}}=1.01(5)$.\footnote{The number in parentheses is an estimate of the error. It accounts
for the statistical uncertainty of the fitted data and to how much
the fitted value changes if we increase, shrink, or shift the fitting
region by a few lattice spaces. In all cases, we verify that the reduced
weighted error sum $\chi^{2}\le2$.} For completeness, we report that these values were obtained by fitting
the data within the range $x_{\text{min}}\leq rD\leq x_{\text{max}}$
where $\left(x_{\text{min}},x_{\text{max}}\right)\approx\left(4,10\right)$,
$\left(10,20\right),$ $\left(15,30\right)$, $\left(20,40\right)$,
$\left(25,50\right)$, and $\left(25,60\right)$, for $D=0.5625,\,1.0625,\dots,\,3.0625$,
respectively. 

\begin{figure}[b]
\begin{centering}
\includegraphics[clip,width=0.99\columnwidth]{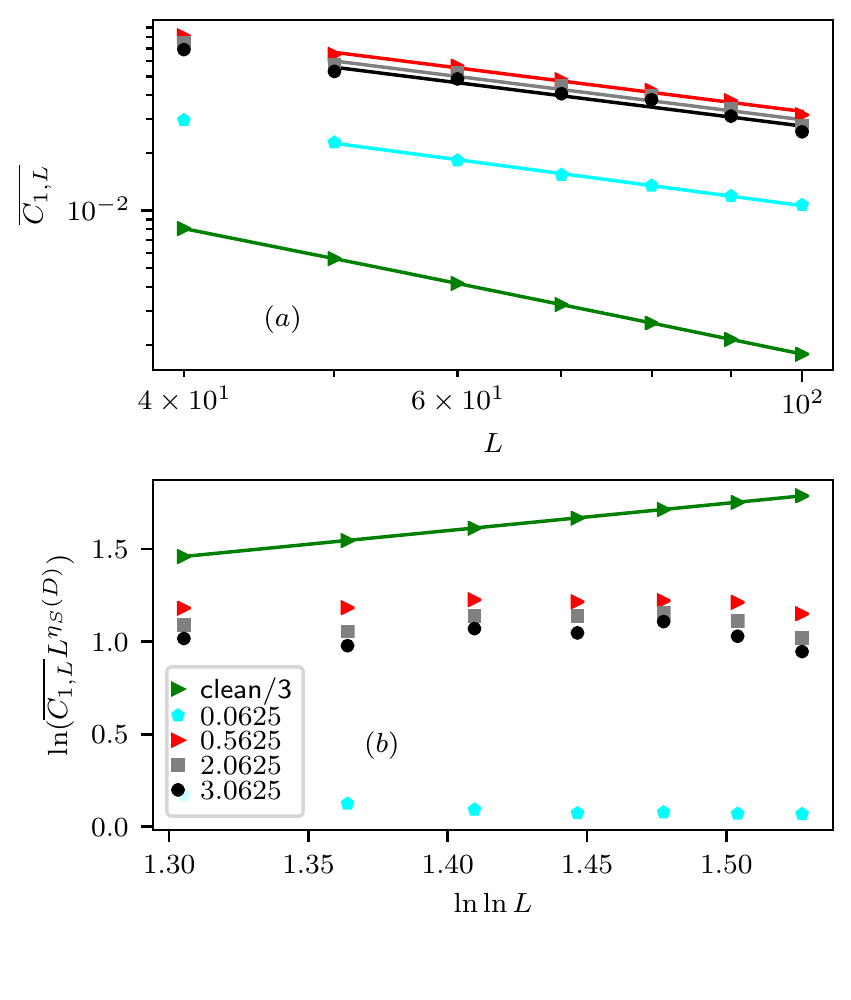}
\par\end{centering}
\caption{(a) The average end-to-end correlation $\overline{C_{1,L}}=-\overline{\left\langle \mathbf{S}_{1}\cdot\mathbf{S}_{L}\right\rangle }$
of the AF Heisenberg chain as a function of the system size and for
various values of the disorder parameter $D$. The straight lines
are power-law fits to $aL^{-\eta_{s}}$ from which we obtain $\eta_{s}=1.02(7)$
for all values of $D$ except for $D=1/16$, in which $\eta_{s}=1.1$.
In the clean case, however, the solid line is the fit to Eq.~\eqref{eq:CL}
already reported in Fig.~\ref{fig:CL}. (b) A replot of data in (a)
with $\overline{C_{1,L}}$ multiplied by $L^{\eta_{s}(D)}$, with
$\eta_{s}(D)=1+\delta_{D,0}$. The solid line is the same in Fig.~\hyperref[fig:CL]{\ref{fig:CL}(b)}.
For an easier comparison, in panel (b) we have divided $C_{1,L}$
by a factor of 3 in the clean case.\label{fig:C1L}}
\end{figure}

We now study the mean end-to-end correlation $\overline{C_{1,L}}=-\overline{\left\langle \mathbf{S}_{1}\cdot\mathbf{S}_{L}\right\rangle }$
for even $L$. In Fig.~\hyperref[fig:C1L]{\ref{fig:C1L}(a)}, $\overline{C_{1,L}}$
is plotted as a function of the system size $L$ for various values
of the disorder parameter $D$, including $D=0$ (the clean system)
for comparison. Disregarding logarithmic corrections, $C_{1,L}\sim L^{-\eta_{s}(0)}$,
with a clean surface exponent $\eta_{s}(0)=2$ that is greater than
the bulk exponent $\eta=1$ (see Fig.~\ref{fig:CL}). In the disordered
case $D\neq0$, however, $C_{1,L}\sim L^{-\eta_{s}(0)}$, with surface
exponent $\eta_{s}(D\neq0)=1$~\citep{YoungFisher_1998_End2EndRTFI},
which is less than the bulk one $\eta(D\neq0)=2$. In the SDRG framework,
$C_{1,L}$ is proportional to the probability that the first and last
spins form a singlet. Thus, on average, it decays with the system
$\sim L^{-1}$. Our data (see Fig.~\ref{fig:CL}) are clearly compatible
with this prediction. We notice a nonmonotonic behavior of $C_{1.L}$
as a function of $D$. It increases from $D=0$ up to $D^{*}\approx0.7(2)$
and diminishes for larger $D$. A similar behavior was also found
in the free-fermion case $\Delta=0$ for the longitudinal correlation
$\overline{C_{1,2}^{z}}$~\citep{Getelina_CA}. 

In order to highlight a possible logarithmic factor, we replot in
Fig.~\hyperref[fig:C1L]{\ref{fig:C1L}(b)} the data from Fig.~\hyperref[fig:C1L]{\ref{fig:C1L}(a)}
with $C_{1,L}$ multiplied by $L^{\eta_{s}(D)}$, with $\eta_{s}(D)=1+\delta_{D,0}$.
In the clean case, $C_{1,L}$ clearly has a logarithmic multiplicative
factor {[}as already reported in Fig.~\hyperref[fig:CL]{\ref{fig:CL}(b)}{]}.
In the disordered case, our data are compatible with its absence.
Even for the smallest value of disorder $D=1/16$, the corresponding
value of $\eta_{s}\approx1.1$ is already far from the clean value
$2$. This indicates that the end-to-end correlation is less affected
by the clean-dirty crossover when compared to the bulk correlations.
More interestingly, the logarithmic factor (if any) is strongly affected
by disorder indicating its absence.

\begin{figure}[tb]
\begin{centering}
\includegraphics[clip,width=0.99\columnwidth]{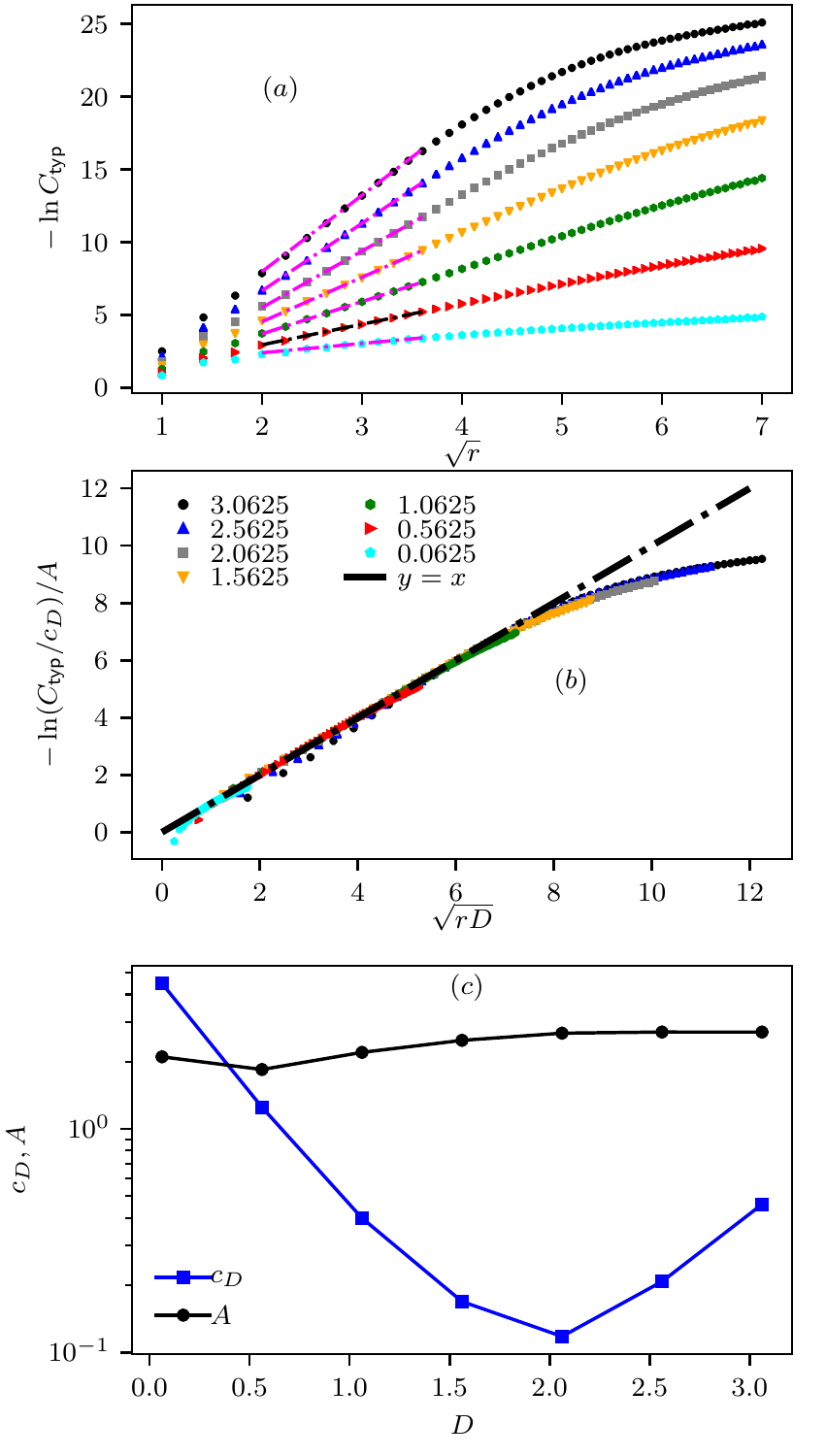}
\par\end{centering}
\caption{(a) The typical correlation function \eqref{eq:Ctyp} of the AF Heisenberg
chain for $L=100$ and different values of the disorder strength $D$.
The magenta lines are the best fits to Eq.~\eqref{eq:C-typ}, from
which the data collapse in (b) is produced. (c) The values of the
fitting parameters $A$ and $c_{D}$ as a function of the disorder
parameter $D$.\label{fig:Ctyp} }
\end{figure}

We now turn our attention to the typical value of the correlation
function, 
\begin{equation}
C_{\text{typ}}=\exp\left(\overline{\frac{\sum_{i=L/4}^{3L/4-r}\ln\left|\langle\mathbf{S}_{i}\cdot\mathbf{S}_{i+r}\rangle\right|}{L/2-r}}\right),\label{eq:Ctyp}
\end{equation}
 which is defined analogously to Eq.~\eqref{eq:C}. This quantity
is plotted in Fig.~\ref{fig:Ctyp} as a function of the spin-spin
separation $r$ for $L=100$ and various values of $D$. Figure \hyperref[fig:Ctyp]{\ref{fig:Ctyp}(a)}
shows $\ln C_{\text{typ}}$ vs $\sqrt{r}$, from which the linear
behavior \eqref{eq:C-typ} is confirmed for $\xi_{D}\ll r\ll L$. 

Analogously to the average value (see Fig.~\ref{fig:Crand}), we
produce a data collapse based on Eq.~\eqref{eq:C-typ}. This is done
by fitting Eq.~\eqref{eq:C-typ} to the data in Fig.~\hyperref[fig:Ctyp]{\ref{fig:Ctyp}(a)}
in a region which, presumably, is weakly affected by finite size (see
magenta lines). The resulting collapsed data are shown in Fig.~\hyperref[fig:Ctyp]{\ref{fig:Ctyp}(b)}.
The fitting parameters $A$ and $c_{D}$ are shown in Fig.~\hyperref[fig:Ctyp]{\ref{fig:Ctyp}(c)}
as a function of the disorder parameter $D$. In agreement with Eq.~\eqref{eq:C-typ},
$A$ is disorder independent for large $D$. We attribute the weak
$D$ dependence to the large crossover length in the weak disorder
limit. Similar behavior was found in the free-fermion case $\Delta=0$~\citep{Getelina_CA}.

\begin{figure}[tb]
\begin{centering}
\includegraphics[clip,width=0.99\columnwidth]{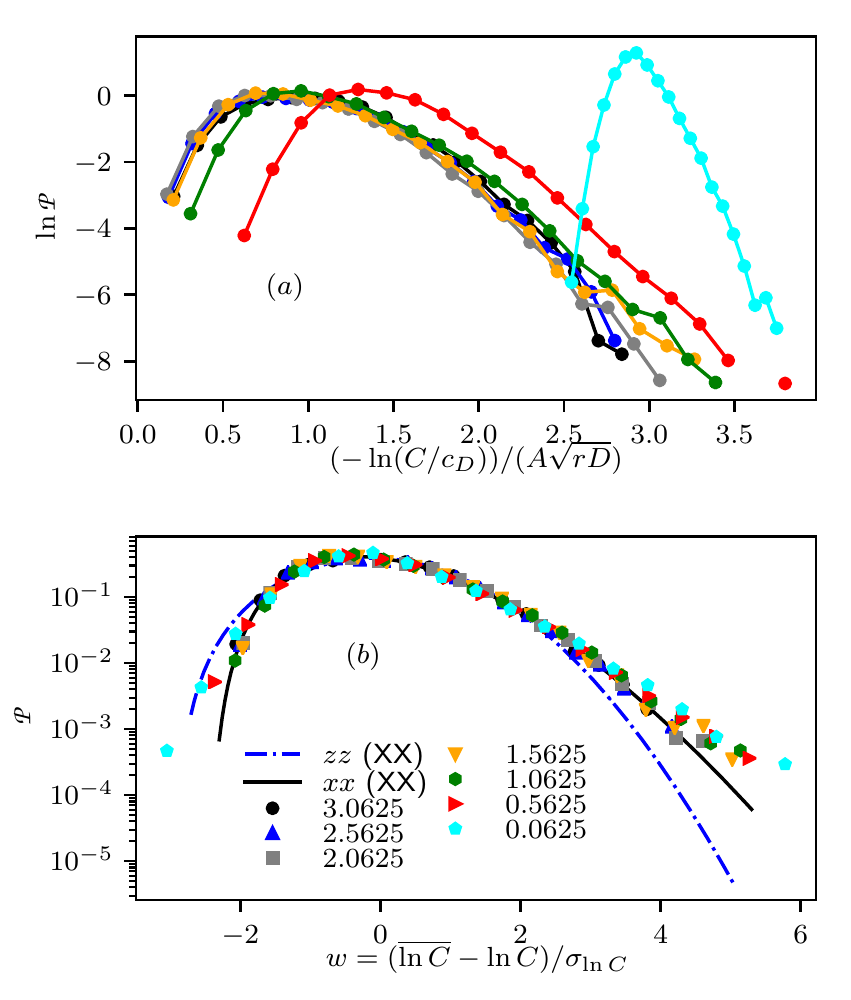}
\par\end{centering}
\caption{Normalized histogram of the correlation function $C=\left|\left\langle \mathbf{S}_{L/4}\cdot\mathbf{S}_{3L/4}\right\rangle \right|$.
In (a), it is rescaled by the parameters $A$ and $c_{D}$ {[}see
Eq.~\eqref{eq:C-typ}{]} given in Fig.~\hyperref[fig:Ctyp]{\ref{fig:Ctyp}(c)}.
(b) shows the same data rescaled by its average and standard deviation.
The solid black line is a fit to Eq.~\eqref{eq:Pfit} (see text).
The histogram was built using $2\times10^{4}$ distinct disorder configurations
of coupling constants $\left\{ J_{i}\right\} $. The blue dashed and
black solid lines are, respectively, the associated distributions
for the XX chain {[}$\Delta=0$ in \eqref{eq:C-mean}{]} for the longitudinal
$C^{z}$ and transverse $C^{x}$ spin-spin correlations. \label{fig:Pdist}}
\end{figure}

Finally, we now study the distribution of spin-spin correlations.
We restrain ourselves to the quantity $C=\left|\left\langle \mathbf{S}_{L/4}\cdot\mathbf{S}_{3L/4}\right\rangle \right|$
for chains of size $L=100$ and many values of $D$. In Fig.~\hyperref[fig:Pdist]{\ref{fig:Pdist}(a)}
we plot the distribution of $-\ln\left(C/c_{D}\right)/\left(A\sqrt{\gamma_{D}r}\right)$,
with $r=L/2$ and $A$ and $c_{D}$ being the fitted values in Fig.~\hyperref[fig:Ctyp]{\ref{fig:Ctyp}(c)}.
The data collapse for the largest values of $D$ confirms the conjecture
of Ref.~\citealp{Fisher_1994_RTFI} which states that the distribution
of $-\ln\left|C_{i,j}\right|/\sqrt{\left|i-j\right|}$ converges to
a nontrivial distribution for large spin-spin separation $\left|i-j\right|\gg1$.
Here, in addition, we conclude that $\ln\left|C_{i,j}\right|/\sqrt{\gamma_{D}\left|i-j\right|}$
converges to a nontrivial, narrow, and universal (disorder-independent)
distribution for $\gamma_{D}\left|i-j\right|\gg1$. The same observation
was reported in the free-fermion case $\Delta=0$~\citep{Getelina_CA}.

We are now interested in the functional form of this nontrivial distribution.
We thus study the distribution of $w=\frac{\overline{\ln C}-\ln C}{\sigma_{\ln C}}$
(with $\sigma_{x}^{2}=\overline{x^{2}}-\overline{x}^{2}$ being the
variance of $x$), which is shown in Fig.~\hyperref[fig:Pdist]{\ref{fig:Pdist}(b)}. 

In Ref.~\citealp{Getelina_CA}, the distributions of the transverse
($C^{x}$) and longitudinal ($C^{z}$) correlations for the XX model
{[}$\Delta=0$ in Eq.~\eqref{eq:HXXZ}{]} were shown to be well fitted
by 

\begin{equation}
{\cal P}(w)=B\exp\left\{ -\left|\frac{w-w_{1}}{\delta_{1}}\right|^{\gamma_{1}}-\left(\frac{\delta_{2}}{w-w_{2}}\right)^{\gamma_{2}}\right\} .\label{eq:Pfit}
\end{equation}
 The first term in the exponential dictates the weak-correlation behavior
$w\gg1$, which, naively, is expected to be a Gaussian; that is, $\gamma_{1}$
is expected to be $2$. Thus, $w_{1}$ and $\delta_{1}$ are, respectively,
the associated mean and width. The second term in the exponential
dictates the strong-correlation regime. A sharp cutoff, represented
by $\omega_{2}$, is expected since the correlations $C_{i,j}$ cannot
be arbitrarily large in absolute value. Thus, $w>w_{2}$. The parameters
$\delta_{2}$ and $\gamma_{2}$ are the associated width and exponent,
respectively. The parameter $B$ is just the normalization. The fitted
values in that work for the transverse correlations are $\delta_{1}=1.66$,
$\delta_{2}=79$, $w_{1}=-1.45$, $w_{2}=-2.51$, $\gamma_{1}=1.71$,
and $\gamma_{2}=0.41$, which is plotted as a black solid line in
Fig.~\hyperref[fig:Pdist]{\ref{fig:Pdist}(b)}. Surprisingly, it
fits our data quite satisfactorily. It is thus tempting to conjecture
that the distribution of the transverse correlation in the model Hamiltonian
\eqref{eq:HXXZ} is $\Delta$ independent in the infinite-randomness
regime $-\frac{1}{2}<\Delta\leq1$. For comparison, we also plot the
distribution of the longitudinal correlations $C^{z}$ (blue dashed
line) of the XX model obtained in Ref.~\citealp{Getelina_CA}. 

Thus, we conclude that the distribution of $\ln\left|C_{i,j}^{x}\right|/\sqrt{\gamma_{D}\left|i-j\right|}$
in the long-distance regime $\gamma_{D}\left|i-j\right|\gg1$ converges
to a nontrivial distribution which is narrow, universal (disorder
independent), and possibly $\Delta$ independent. The same conclusions
apply for the distribution of longitudinal correlations, except that
it is $\Delta$ dependent.

\section{Conclusions and discussion\label{sec:Conclusion}}

In this work, we have studied various measures of the ground-state
spin-spin correlations $C_{i,j}=\left\langle \mathbf{S}_{i}\cdot\mathbf{S}_{j}\right\rangle $
of the AF spin-$1/2$ Heisenberg chain {[}$\Delta=1$ in Eq.~\eqref{eq:HXXZ}{]}
with random coupling constants. We applied the recently developed~\citep{Xavier_ADMRG}
adaptive strategy, which enabled us to study strongly (quenched) disordered
systems using the unbiased DMRG method.

Our data are entirely compatible with the SDRG analytical predictions~\citep{Fisher_1994_RTFI,CARandomSpinChain_Hoyos2007}.
Specifically, regarding the bulk correlations \eqref{eq:C-mean} in
the regime $1\ll\gamma_{D}\left|i-j\right|\ll L$, we verified that
the exponent $\eta=2$ and prefactor difference $c_{\text{o}}-c_{\text{e}}=1$
are universal (disorder independent). Our data confirm that the typical
value of the correlations {[}Eq.~\eqref{eq:C-typ}{]} decay stretched
exponentially with the spin-spin separation with universal exponent
$\psi=1/2$. Furthermore, we have confirmed the observation of Ref.~\citealp{Florencie_XXZCrossLength}
that the relevant length scale is the inverse Lyapunov exponent $\gamma_{D}$
in Eq.~\eqref{eq:Lyapunov}, which plays the role of the clean-dirty
crossover length. This observation was made precise in the XX chain
case ($\Delta=0$). In that case, this length scale is the inverse
of the Lyapunov exponent of a single-parameter theory of the associated
free-fermion system with particle-hole symmetry~\citep{mard-etal-prb14}.
We have also studied the distribution of the spin-spin correlations
for a fixed distance and confirmed the conjecture of Ref.~\citealp{Fisher_1994_RTFI}
that $\ln\left|C_{i,j}\right|/\sqrt{\left|i-j\right|}$ converges
to a nontrivial distribution for $\left|i-j\right|\gg1$. We have
also studied the mean value of the end-to-end correlations $C_{1,L}$
and confirmed that it decays $\sim L^{-\eta_{s}}$ with universal
surface exponent $\eta_{s}=1$~\citep{YoungFisher_1998_End2EndRTFI}.
All these results were thoroughly confirmed by many others using different
methods~\citep{Igloi_2005_SDRGREview,Igloi_2018_SDRGREview}. 

Let us now summarize our new findings. The first one is in regard
to the end-to-end correlations on the clean system. Our data are compatible
with the predicted surface exponent $\eta_{s}=2$. In addition, we
showed the existence of a logarithmic correction with effective exponent
$\theta=1.5(1)$ {[}see Eq.~\eqref{eq:CL}{]}. In the presence of
disorder, this logarithm correction disappears. With respect to the
distribution of correlations, we have found that the distribution
of $\ln\left|C_{i,j}\right|/\sqrt{\gamma_{D}\left|i-j\right|}$ converges
to a nontrivial, narrow, and universal distribution for $\gamma_{D}\left|i-j\right|\gg1$.
We have also found that, within our statistical precision, it is equal
to that of the transverse correlations of the XX chain reported in
Ref.~\citep{Getelina_CA}. It is thus tempting to conjecture that,
besides being nontrivial, narrow, and universal, the distribution
of $\ln\left|C_{i,j}^{x}\right|/\sqrt{\gamma_{D}\left|i-j\right|}$
does not depend (or depends weakly) on $\Delta$.

One reported result that we have not confirmed is the logarithmic
factor on the mean correlations of the disordered chain~\citep{XXXLogCorr_Sandvik2016}.
As we have shown, our data are compatible with its absence in both
the mean (see Figs.~\ref{fig:Crand} and \ref{fig:eta-c}) and typical
(see Fig.~\ref{fig:Ctyp}) values of the bulk correlations, as well
as in the mean value of the end-to-end correlation (see Fig.~\ref{fig:C1L}).
Evidently, we cannot exclude (although it very implausible) a logarithmic
factor appearing for system sizes larger than the ones studied here.
If that is the case, we recall that the adaptive DMRG method employed
here starts with the near-clean wave function, which does have a logarithmic
factor in its two-point correlation. The fact that we do not detect
it when the disorder strength is increased strongly suggests that
the origin of the logarithmic factor, if one exists, is unrelated
to that of the clean system.

Currently, it is unclear why the zero-temperature quantum Monte Carlo
study of Ref.~\citealp{XXXLogCorr_Sandvik2016} predicts a logarithmic
correction. The only suggestion that comes to us is finite-size effects.
As shown in Fig.~\hyperref[fig:Crand]{\ref{fig:Crand}(c)}, the finite-size
corrections are still strong for $D\approx2$ even for system sizes
$L\approx100$. More importantly, the finite-size correction promotes
a slow increase in the correlations which can be interpreted as a
logarithmic correction~\citep{Getelina_CA}. Interestingly, $D=2$
is the strongest disorder parameter value studied in Ref.~\citealp{XXXLogCorr_Sandvik2016}.
However, those authors considered periodic boundary conditions where
finite-size effects are presumably smaller. In addition, they were
able to study chains with sizes larger than ours.

Finally, we would like to point out that logarithmic factors are predicted
by the SDRG method. They appear in the susceptibility and specific
heat of infinite-randomness critical chains (but not in the correlations)~\citep{Fisher_1994_RTFI}
and in certain quantities of critical chains at a Kosterlitz-Thouless
like transition~\citep{altman-etal-prl04,vojta-etal-jp11,juhasz-kovacs-igloi-epl14}.
Interestingly, logarithmic factors appear in the correlations of the
clean Heisenberg chain ($\Delta=1$) and in the weakly disordered
XXZ chain at the point $\Delta=-\frac{1}{2}$. In both cases, the
associated renormalization-group flow is of the Kosterlitz-Thouless
type~\citep{ristivojevic-etal-npb12}.

In conclusion, our numerical results are in agreement with those predicted
by the SDRG method, which, presumably, yields asymptotically exact
results for the ground-state properties of the model Hamiltonian \eqref{eq:HXXZ}
in the parameter region $-\frac{1}{2}<\Delta\leq1$. We have also
shown that the adaptive DMRG method is capable of tackling one-dimensional
disordered systems. It can be easily implemented using the standard
DMRG method without much more coding effort, and therefore, it adds
to the toolbox of unbiased theoretical methods for disordered systems.
\begin{acknowledgments}
We thank F. Alcaraz, R. Pereira, A. Sandvik, R. Juh\'asz, and N.
Laflorencie for useful discussions. We acknowledge the financial support
of the Brazilian agencies FAPESP and CNPq.


%
\end{acknowledgments}

\bibliographystyle{apsrev4-1}
\bibliography{HeisDMRG}

\end{document}